\documentclass{jps-cp}
\usepackage{txfonts} 
\usepackage{units}	
\usepackage[squaren,Gray]{SIunits} 
\usepackage[arrowdel]{physics}
\usepackage{color}

\title{Unconventional charge density wave order in ErTe$_{3}$: symmetry breakings and fluctuations probed by Raman scattering}

\author{Théotime \textsc{Freitas}$^{1}$, Mattia \textsc{Udina}$^{1,2}$, Alexandr \textsc{Alekhin}$^{1}$, Niloufar \textsc{Nilforoushan}$^1$, Sarah \textsc{Houver}$^1$, Alain \textsc{Sacuto}$^1$, Benito \textsc{A. Gonzalez}$^{3,4,5}$, Ian \textsc{R. Fisher}$^{3,4,5}$, Indranil \textsc{Paul}$^{1}$ and Yann \textsc{Gallais}$^{1,6}$}

\inst{$^{1}$Université Paris Cité, CNRS, Laboratoire Matériaux et Phénomènes Quantiques, Paris, France \\
$^{2}$Institut de Physique et Chimie des Mat\'{e}riaux de Strasbourg (UMR 7504),  Universit\'{e} de Strasbourg and CNRS, Strasbourg, 67200, France \\
$^{3}$Department of Applied Physics, Stanford University, Stanford, CA, USA \\ 
$^{4}$Geballe Laboratory for Advanced Materials, Stanford University, Stanford, California 94305, USA \\ 
$^{5}$Stanford Institute for Materials and Energy Sciences, SLAC, Menlo Park, California 94025, USA
$^{6}$Institut Universitaire de France (IUF)}

\email{theotime.freitas@u-pariscite.fr}
\email{yann.gallais@u-pariscite.fr}

\recdate{September 24, 2026}

\abst{Recent experiments have revealed unexpected mirror-symmetry breakings in the charge density-wave (CDW) state of rare-earth tritellurides, suggesting a putative ferro-axial CDW state. Here we report polarization-resolved Raman scattering measurements tracking the point-group symmetry breakings across the CDW phase transition in the rare-earth tritelluride ErTe$_3$. We show that the mirror-symmetry breakings follow closely the emergence of the CDW order, suggesting a single component CDW order parameter. Furthermore, Raman measurements in the $A_{2g}$ symmetry channel above the CDW transition temperature indicate the absence of significant ferro-axial fluctuations. We discuss the broad implications of these findings with respect to the nature of the CDW state in the tritellurides.}

\kword{Raman scattering, charge density wave, ErTe$_{3}$, ferro-axial order, amplitude mode}

\begin{document}
\maketitle
The rare-earth tri-telluride (ReTe$_3$, RE: rare earth) are prototypical weakly correlated 2D charge density wave (CDW) materials where the CDW order is thought to be mostly driven by Fermi surface nesting \cite{ru_effect_2008, yao_theory_2006,brouet_angle-resolved_2008}. They consist of stacked Te square nets with a glide plane symmetry that weakly breaks the equivalence between $a$ and $c$ in-plane axis (here we use standard orthorhombic notations for the labelling the axis). They exhibit a single $Q$ CDW transition with a CDW wave-vector $Q_{CDW}$ along the in-plane $c$ axis. For heavy rare-earth elements there is an additional second phase transition at lower temperatures with a CDW ordering along the orthogonal $a$ axis. While initially thought to be well unterstood in terms of a Fermi surface nesting instability, the exact nature of the CDW order of ReTe$_3$ was recently put into question by several measurements indicating a monoclinic rather than orthorhombic point group in the CDW state \cite{wang_axial_2021,singh_ferroaxial_2025,wulferding_magnetic_2025,Freitas2026,Jack2026}. Specifically, Raman scattering, second-harmonic generation and electron diffraction measurements indicate the breaking of the vertical mirror reflection symmetries of the Te square in the CDW state. The observation of a monoclinic CDW state has been interpreted as the consequence of a ferro-axial CDW order. A theoretical description of this ferro-axial CDW state in terms of a multi-component order parameter involving Te atoms orbital degrees of freedom was proposed to explain the experimental observations \cite{alekseev_charge_2024}. However, a multi-component order parameter with invidual components belonging to different irreducible representations should result into a split phase transition \cite{Freitas2026,Kivelson2026}. This seems at odd with previous X-ray, thermodynamic and transport measurements which suggest a single continuous second-order like phase transition at $T_{CDW}$ \cite{ru_effect_2008}. An alternative scenario, proposing a single component CDW order parameter but with a $Q_{CDW}$ tilted away from the cristallographic c-axis that would break all mirror symmetries, was shown to reproduce the Raman data and their strain dependence \cite{udina_antisymmetric_2025,Freitas2026}. The proposed tilt is small (less than 0.5 degree) and may have been overlooked in previous diffraction experiments. In addition, it was also recently proposed that the monoclinic distortion might the consequence of an enhanced susceptibility of the CDW state to residual symmetry-breaking strains \cite{Kivelson2026}. 
\par
Here we address the issue of the nature of the CDW order in ReTe$_3$ by tracking the different symmetry breakings and their associated fluctuations as a function of temperature close to the CDW phase transition. For this we use polarization-resolved Raman scattering as a function of temperature in the compound ErTe$_3$. Our data show that the different symmetry-breakings track each other across the CDW phase transition, confirming the presence of only a single phase transition in agreement with a single component CDW order parameter. Besides, symmetry-resolved Raman spectra above the $T_{CDW}$ indicate the absence of any significant enhancement of fluctuations related to ferro-axial or nematic instabilities associated with CDW order. Overall, our data are consistent with a monoclinic distortion being at best a secondary instability of the CDW order.


Single crystals of ErTe$_{3}$ were grown by the self-flux method using excess tellurium and characterized by X-ray diffraction. They display a first CDW transition towards the single Q CDW state at $T_{CDW}\sim$ \unit{265}{\kelvin} and a second transition toward a double Q CDW state at $\sim$ \unit{155}{\kelvin}. Here, we will discuss only data  above 160K, near the first CDW transition \cite{ru_effect_2008}. Raman scattering experiments were carried out on two different single crystals (sample A and B) and in different scattering geometries, both yielding similar results. The crystal was exfoliated to obtain fresh surfaces prior to each measurements. A \unit{532}{\nano\meter} solid state laser was used as the excitation source. The imaginary part of the Raman response function $\chi''(\omega)$ was extracted from the raw spectra by correcting for the Bose factor \cite{Freitas2026}.

\begin{figure}[h]
\centering
\includegraphics[width=0.99\linewidth]{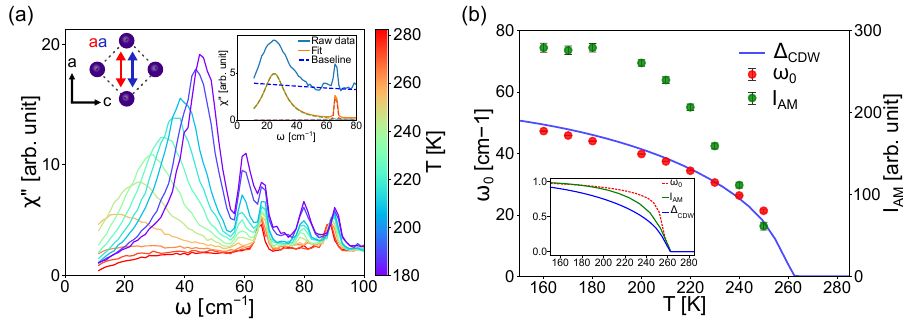}
\caption{(a) : Temperature dependence of the low energy Raman response for ErTe$_{3}$ in the $aa$ polarization configuration (see left inset). The right inset shows an example of fit of the amplitude mode using a damped oscillator function. The color bar indiates the temperature values. (b) : Temperature dependence of the energy $\omega_0$ and intensity $I_{AM}$ of the amplitude mode. The energy and the intensity were obtained by using the fit above. The theoretical temperature dependence of the CDW gap $\Delta_{CDW}$ within a weak-coupling description of the CDW is overlaid in blue. It was rescaled to fit the temperature dependence of $\omega_0$. The inset shows the theoretical dependence of $\Delta_{CDW}$, $I_{AM}$ and $\omega_0$ in a weak-coupling description of the CDW for perfect nesting \cite{cea_nature_2014}. The calculations are performed in the limit $\frac{\Delta_{CDW}}{\omega_0} \gg 1$ which is relevant for ReTe$_3$ given the energy of the CDW gaps extracted from previous Raman and ARPES measurements \cite{eiter_alternative_2013,brouet_angle-resolved_2008}.} 
\label{fig1}
\end{figure}

Figure \ref{fig1}.(a) show the low energy temperature dependent Raman response of ErTe$_{3}$ sample A across $T_{CDW}\sim \unit{265}{\kelvin}$ using the $aa$ polarization configuration (incoming and outgoing photon polarisation along the crystallographic $a$-axis, see inset). The overall temperature of the spectrum agrees with previously published Raman data on ErTe$_3$ \cite{eiter_alternative_2013} and other ReTe$_3$ \cite{lavagnini_raman_2010}. A characteristic feature of the CDW order in the Raman spectrum is the emergence of the CDW amplitude mode (AM) below $T_{CDW}$. The energy of the mode is around $\sim$ \unit{50}{\centi\rp\meter} at \unit{180}{\kelvin} and progressively softens as $T_{CDW}$ is approached from below as expected for a second-order phase transition. We also observe several additional phonon modes (around $\sim$ \unit{60}{\centi\rp\meter} and $\sim$ \unit{80}{\centi\rp\meter}) which do not show any appreciable softening. They are assigned to zone-folded optical phonon which are weakly coupled to the CDW but become Raman-active below $T_{CDW}$ due to the periodic lattice distortion. Focusing on the AM, Fig. \ref{fig1}.(b) shows the temperature dependence of the energy $\omega_{0}$ and the normalized intensity $I_{AM}$ of the AM obtained by fitting it with a damped oscillator function (see inset). Both the intensity and the energy of the AM decrease continuously close to $T_{CDW}$. The AM energy $\omega_0$ follows qualitatively the temperature dependence of the CDW gap $\Delta_{CDW}$ as obtained from a BCS-like weak-coupling theory of the CDW state \cite{cea_nature_2014} while the AM intensity $I_{AM}$ displays a somewhat faster increase below $T_{CDW}$. In the inset we show the theoretical temperature dependence of $\omega_0$ and $I_{AM}$ within a weak-coupling and perfect nesting scenario of the CDW instability of ErTe$_3$ \cite{cea_nature_2014,udina_theory_2019}. Both quantites display a much steeper temperature dependence than $\Delta_{CDW}$ which, while qualitatively consistent with the experiment for $I_{AM}$, does not match the behaviour of $\omega_0$. This discrepancy likely highlights the deficiency of a simple Peierls-like perfect nesting description of the CDW state even in a weakly correlated system like ReTe$_3$ \cite{johannes_fermi_2008}. A more realistic theoretical description of the temperature dependence of the AM likely requires to take into account wave-vector depedendent electron-phonon interaction as well as imperfect nesting \cite{eiter_alternative_2013,maschek_wave-vector-dependent_2015}.
  
\begin{figure}[t]
\centering
\includegraphics[width=0.99\linewidth]{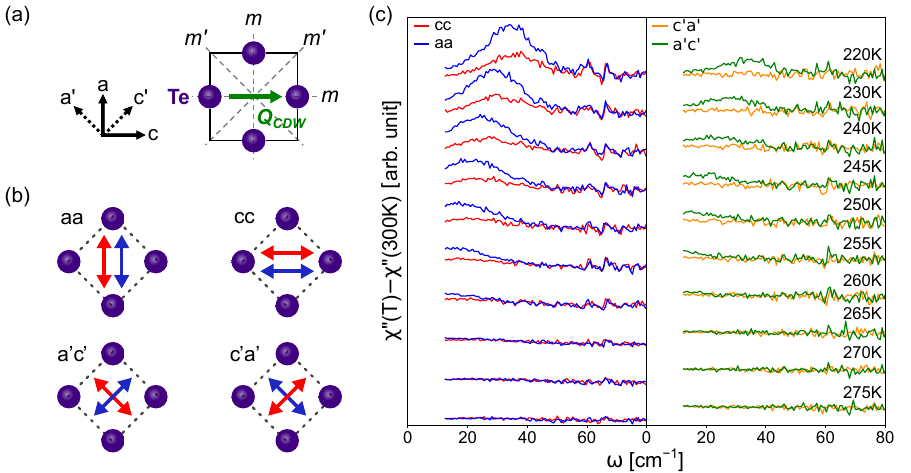}
\caption{(a) : Definition of the axis $a$, $c$, $a'$ and $c'$. Sketch of the Te square showing the mirror reflection symmetries $m$ and $m'$, and $Q_{CDW}$ (b) : Polarization configurations showing the incoming (red) and outgoing (blue) photon polarizations with respect to the Te square plane. (c) : Temperature dependence of the low energy Raman response for ErTe$_{3}$ subtracted by the \unit{300}{\kelvin} Raman response in $aa$, $cc$, $a'c'$ and $c'a'$ polarization configuration.}
\label{fig2}
\end{figure}

We now focus on the polarization dependence of the AM across the CDW transition as a marker of the different symmetry breakings. The mirror reflection symmetries of the Te square plane along with $Q_{CDW}$ are depicted in figure \ref{fig2}(a). In Fig. \ref{fig2}.(b) the four different polarization configurations measured are shown, with the red (blue) arrows indicating the polarization of the incoming (outgoing) photons. Figure \ref{fig2}.(c) shows the temperature dependent Raman response of ErTe$_{3}$ (sample B) after subtraction of the normal-state Raman response at \unit{300}{\kelvin}, for the four polarization configurations described before. For temperature above the transition, the spectra display a featureless electronic continuum typical of a metallic state in all the polarization configurations. Below $T_{CDW}$ the AM emerges in both parallel and crossed polarization configurations. However it is evident from the data that by contrast with the spectra above $T_{CDW}$, the AM intensity is strongly polarisation dependent, reflecting distinct symmetry breakings in the CDW state. For parallel polarization the AM anisotropy, with $I_{AM,aa}> I_{AM,cc}$, simply reflects the $m'$ mirror symmetry breaking associated to the uni-directional CDW ordering with $Q_{CDW}$ // c. We note here that while $m'$ mirror reflection is already formally broken above $T_{CDW}$ by the glide-plane symmetry, the spectra above $T_{CDW}$ do not display any appreciable fingerprint of this weak structural symmetry breaking: the spectra in $aa$ and $cc$ geometry are essentially identical above $T_{CDW}$. For cross polarization however, the difference between $a'c'$ and $c'a'$ is incompatible with an orthorhombic point group. It indicates an additional $m$ mirror symmetry breaking that cannot be linked to $Q_{CDW}$ if it lies along the $c$ direction. 

\begin{figure}[t]
\centering
\includegraphics[width=0.99\linewidth]{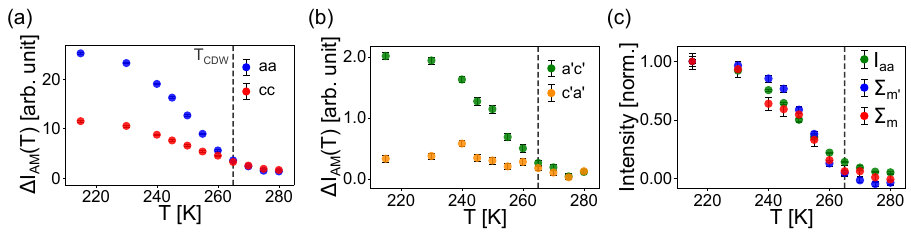}
\caption{(a-b) : Temperature dependence of the integrated area of the AM subtracted by the area at \unit{300}{\kelvin} for parallel (a) and crossed polarization configurations (b). (c) : Temperature dependence of the normalized integrated intensity of $I_{aa}$, $\Sigma_{m}$ and $\Sigma_{m'}$.}
\label{fig3}
\end{figure}

The evolution of the polarization-resolved low energy Raman response as a function of temperature was monitored by integrating the Raman response up to \unit{57}{\centi\rp\meter}, a spectral range which includes the AM mode at all investigated temperatures. In order to highlight the changes across $T_{CDW}$ the integrated area were subtracted by the normal-state response at \unit{300}{\kelvin}. The results are shown in Fig. \ref{fig3} for parallel (a) and crossed (b) polarizations configurations. In both parallel $aa$/$cc$ and crossed $a'c'$/$c'a'$ configurations the integrated quantities are weakly temperature dependent above $T_{CDW}$ (this weak temperature dependence is further discussed below), but display a pronounced enhancement and a clear differentiation below $T_{CDW}\sim$ 265K.  To further quantify the temperature evolution of the different symmetry breakings across $T_{CDW}$, we introduce the symmetry-resolved quantities $\Sigma_{m'}$ and $\Sigma_{m}$, which serve as markers of the corresponding broken mirror symmetries. We define \cite{Freitas2026}: $\Sigma_{m'}=\frac{I_{aa}-I_{cc}}{I_{aa}+I_{cc}}$ and $\Sigma_{m}=\frac{I_{a'c'}-I_{c'a'}}{I_{a'c'}+I_{c'a'}}$
where $ I_{ij}$ corresponds to the integrated area of the Raman response for $ij$ polarization. Note that $\Sigma_{m'}$ ($\Sigma_m$) changes sign upon $m'$ ($m$) mirror reflection and is thus a marker of $m'$ ($m$) mirror symmetry breaking. In Fig. \ref{fig3}.(c) are displayed the temperature dependence of $\Sigma_{m'}$ and $\Sigma_m$ along with $I_{aa}$ which tracks the CDW translational symmetry breaking as discussed above. All three quantities are subtracted by their \unit{300}{\kelvin} value and normalized to their $T=\unit{210}{\kelvin}$ values. Within the accuracy of our measurements they exhibit very similar temperature dependences with relatively sharp upturns below $T_{CDW}$. Our data are thus consistent with a single phase transition at $T_{CDW}$ where translational and mirror symmetries are simultaneously broken. While we cannot rule out a small ($\leq$ \unit{5}{\kelvin}) split phase transition as expected for a multi-component CDW order parameter scenario, the present data appear more in-line with a single component CDW order parameter \cite{udina_antisymmetric_2025,Freitas2026}.

\par
\begin{figure}[t]
\centering
\includegraphics[width=0.99\linewidth]{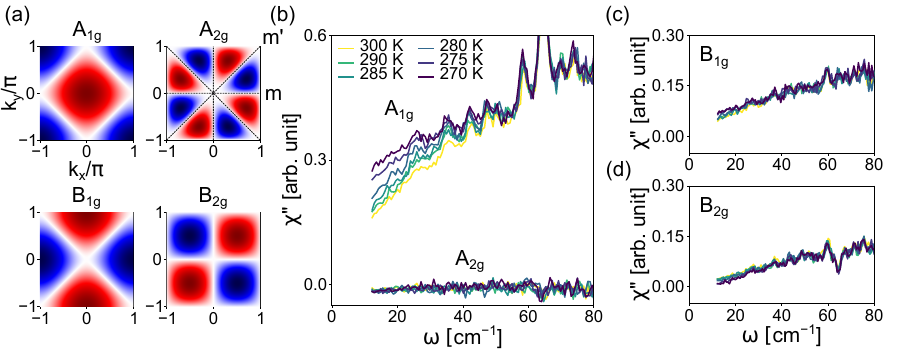}
\caption{(a) : Sketch of the structure of the irrep in k-space. The mirrors $m$ and $m'$ are depicted for $A_{2g}$ irrep to highlight the mirror symmetry breakings. Temperature dependence of the low-energy Raman response decomposed into the irreducible representations of the $D_{4h}$ crystallographic point group. (b) Low energy Raman spectra in the $A_{1g}$ and $A_{2g}$, (c) $B_{1g}$, and (d) $B_{2g}$ irrep above $T_{CDW}$.}
\label{fig4}
\end{figure}

Having discussed the nature of the symmetry-breakings in the CDW ordered phase, we now examine their associated fluctuations above $T_{CDW}$. As demonstrated in the case of nematicity in iron-based superconductors Raman scattering is a relevant symmetry-resolved probe of fluctuations associated to point-group symmetry breakings \cite{Gallais2013}. Neglecting the glide-plane $m'$ symmetry breaking associated with the stacking of Te square net and thus assuming a $D_{4h}$ tetragonal point group, the in-plane Raman response can be decomposed into the $A_{1g}$, $A_{2g}$, $B_{1g}$, and $B_{2g}$ irreducible representations (irrep). The $A_{1g}$ irrep is fully symmetric while the nematic $B_{1g}$ ($B_{2g}$) irrep breaks the $C_4$ and $m'$ ($m$) symmetries. The $A_{2g}$ irrep on the other hand breaks both $m$ and $m'$ while preserving $C_4$ rotational symmetry and is thus directly linked to putative ferro-axial degrees of freedom \cite{singh_ferroaxial_2025} (see figure \ref{fig4}(a) for a depiction of the k-space structure of the irrep). The Raman response for each irrep was extracted from appropriate combinations of the Raman spectra measured in the linear and circular polarization configurations \cite{devereaux_inelastic_2007}. Figure \ref{fig4}(a-c) displays the temperature dependence above $T_{CDW}$ of the low energy Raman response in $A_{1g}$, $A_{2g}$, $B_{1g}$ and $B_{2g}$ channels. Only the fully symmetric $A_{1g}$ channel exhibits a significant temperature dependence with a sizable enhancement of the low-energy Raman response upon approaching $T_{CDW}$. By contrast the $A_{2g}$ response remains nearly temperature independent and close to zero over the investigated energy range. Together with the absence of any significant temperature dependence in the $B_{1g}$ and $B_{2g}$ channels, we conclude that there is no evidence for critical fluctuations associated to nematic/mirror symmetry breakings and/or ferro-axial degrees of freedom. The fluctuations observed in the $A_{1g}$ channel are likely related to point-group symmetry preserving CDW fluctuations with wave-vector along both the nearly degenerate $a$ and $c$ axis \cite{maschek_competing_2018}.

In conclusion temperature dependent Raman measurements across the CDW transition temperature indicate that mirror symmetry breakings track the emergence of the CDW order, ruling out a significant splitting of the phase transition. This finding favours a single rather than multi-component order parameter. In combination with the absence of any significant ferro-axial fluctuations in the $A_{2g}$ above $T_{CDW}$ symmetry channel, the present results suggest that the monoclinic distortion observed in the CDW state may not be linked to a ferro-axial instability.

\par
We acknowledge funding from the Agence National de la Recherche via the grants ANR “SUPER2DTMD” and ANR “Tri-QMat.” This work has been supported by Region Île-de-France in the framework of DIM QuanTiP and DIM SIRTEQ. Crystal growth and characterization at Stanford was supported by the Department of Energy, Office of Basic Energy Sciences, under Contract No. DE-AC02-76SF00515.



\begin{thebibliography}{9}

\bibitem{ru_effect_2008} N. Ru et al. Phys. Rev. B 77, 035114 (2008)
\bibitem{yao_theory_2006} H. Yao, J. A. Robertson, E.-A. Kim and S. A. Kivelson, Phys. Rev. B 74, 245126 (2006)
\bibitem{brouet_angle-resolved_2008} V. Brouet et al., Phys. Rev. B 77, 235104 (2008)
\bibitem{wang_axial_2021} Y. Wang et al., Nature 606, 896–901 (2022)
\bibitem{singh_ferroaxial_2025} B. Singh et al. Nature Physics 21, 1578–1586 (2025)
\bibitem{wulferding_magnetic_2025} D. Wulferding, J. Park, T. Tohyama, S. R. Park and C. Kim, Nature Communications 16, 114 (2025)
\bibitem{Freitas2026} T. Freitas et al. Phys. Rev. Lett. 137, 166502 (2026)
\bibitem{Jack2026} J. Zheng et al. https://arxiv.org/abs/2606.21872 (2026)
\bibitem{maschek_competing_2018} M. Maschek et al. Phys. Rev. B 98, 094304 (2018)
\bibitem{alekseev_charge_2024} S. Alekseev, S. A. A. Ghorashi, R. M. Fernandes and J. Cano, Phys. Rev. B 110, 205103 (2024)
\bibitem{Kivelson2026} S. Zhao, J. May-Mann, I. R. Fisher and S. A. Kivelson, http://arxiv.org/abs/2608.24995 (2026)
\bibitem{udina_antisymmetric_2025}M. Udina and I. Paul, Phys. Rev. Lett. 136, 126505 (2026)
\bibitem{eiter_alternative_2013} H. M. Eiter et al. PNAS 110, 64-69 (2013)
\bibitem{lavagnini_raman_2010} M. Lavagnini et al. Phys. Rev. B 81, 081101 (2010)
\bibitem{cea_nature_2014}  T. Cea and L. Benfatto, Phys. Rev. B 90, 224515 (2014)
\bibitem{udina_theory_2019} M. Udina, T. Cea and L. Benfatto, Phys. Rev. B 100, 165131 (2019)
\bibitem{johannes_fermi_2008} M. Johannes and I. Mazin, Phys. Rev. B 77, 165135 (2008)
\bibitem{maschek_wave-vector-dependent_2015} M. Maschek et al. Phys. Rev. B 91, 235146 (2015)
\bibitem{Gallais2013} Y. Gallais et al. Phys. Rev. Lett. 111, 267001 (2013)
\bibitem{devereaux_inelastic_2007} T. P. Devereaux and R. Hackl, Rev. Mod. Phys. 79, 175 (2007)




\end{thebibliography}
\end{document}